This is an Author Accepted Manuscript version of the following chapter *Nikiforova, A. (2023). Data Security as a Top Priority in the Digital World: Preserve Data Value by Being Proactive and Thinking Security First. In: Visvizi, A., Troisi, O., Grimaldi, M. (eds) Research and Innovation Forum 2022. RIIFORUM 2022. Springer Proceedings in Complexity. Springer, Cham* reproduced with permission of Springer Nature Switzerland AG. The final authenticated version is available online at: https://doi.org/10.1007/978-3-031-19560-0_1

Please cite as:

*Nikiforova, A. (2023). Data Security as a Top Priority in the Digital World: Preserve Data Value by Being Proactive and Thinking Security First. In: Visvizi, A., Troisi, O., Grimaldi, M. (eds) Research and Innovation Forum 2022. RIIFORUM 2022. Springer Proceedings in Complexity. Springer, Cham. https://doi.org/10.1007/978-3-031-19560-0_1*

# Data security as a top priority in the digital world: preserve data value by being proactive and thinking security first

Anastasija Nikiforova [0000-0002-0532-3488]

University of Tartu, Institute of Computer Science, Tartu, 51009, Estonia

Nikiforova.Anastasija@gmail.com

**Abstract**. Today, large amounts of data are being continuously produced, collected, and exchanged between systems. As the number of devices, systems and data produced grows up, the risk of security breaches increases. This is all the more relevant in times of COVID-19, which has affected not only the health and lives of human beings' but also the lifestyle of society, i.e., the digital environment has replaced the physical. This has led to an increase in cyber security threats of various nature. While security breaches and different security protection mechanisms have been widely covered in the literature, the concept of a "primitive" artifact such as data management system seems to have been more neglected by researchers and practitioners. But are data management systems always protected by default? Previous research and regular updates on data leakages suggest that the number and nature of these vulnerabilities are high. It also refers to little or no DBMS protection, especially in case of NoSQL, which are thus vulnerable to attacks. The aim of this paper is to examine whether "traditional" vulnerability registries provide a sufficiently comprehensive view of DBMS security, or they should be intensively and dynamically inspected by DBMS owners by referring to Internet of Things Search Engines moving towards a sustainable and resilient digitized environment. The paper brings attention to this problem and makes the reader think about data security before looking for and introducing more advanced security and protection mechanisms, which, in the absence of the above, may bring no value.

**Keywords**: Data, Database, Internet of Things, Internet of Things Search Engine, IoT, IoTSE, NoSQL, Security, Vulnerability, Oracle, MySQL, PostgreSQL, MongoDB, Redis, IBM Db2, Elasticsearch, SQLite, Cassandra, Memcached, CouchDB, OSINT

## 1 Introduction

Today, in the age of information and Industry 4.0, billions of data sources, including but not limited to interconnected devices (sensors, monitoring devices) forming Cyber-Physical Systems (CPS) and the Internet of Things (IoT) ecosystem, continuously generate, collect, process, and exchange data [1]. With the rapid increase in the number of devices (smart objects or "things", e.g., smartphones, smartwatches, intelligent vehicles etc.) and information systems in use, the amount of data is increasing. Moreover, due to the digitization and variety of data being continuously produced and processed with a reference to Big Data, their value, is also growing and as a result, the risk of security breaches and data leaks, including but not limited to users' privacy [2]. The value of data, however, is dependent on several factors, where data quality and data security that can affect the data quality if the data are accessed and corrupted, are the most vital. Data serve as the basis for decision-making, input for models, forecasts, simulations etc., which can be of high strategical and commercial / business value.

This has become even more relevant in terms of COVID-19 pandemic, when in addition to affecting the health, lives, and lifestyle of billions of citizens globally, making it even more digitized, it has had a significant impact on business [3]. This is especially the case because of challenges companies have faced in maintaining business continuity in this so-called "new normal". However, in addition to those cybersecurity threats that are caused by changes directly related to the pandemic and its consequences, many previously known threats have become even more desirable targets for intruders, hackers. Every year millions of personal records become available online [4-6].

Lallie et al. [3] have compiled statistics on the current state of cybersecurity horizon during the pandemic, which clearly indicate a significant increase of such. As an example, Shi [7] reported a 600% increase in phishing attacks in March 2020, just a few months after the start of the pandemic, when some countries were not even affected. Miles [8], however, reported that in 2021, there was a record-breaking number of data compromises, where "the number of data compromises was up more than 68% when compared to 2020", when LinkedIn was the most exploited brand in phishing attacks, followed by DHL, Google, Microsoft, FedEx, WhatsApp, Amazon, Maersk, AliExpress and Apple. And while [5] suggests that vulnerability landscape is returning to normal, there is another trigger closely related to cybersecurity that is now affecting the world - geopolitical upheaval.

Recent research demonstrated that weak data and database protection in particular is one of the key security threats [4,6,9-11]. This poses a serious security risk, especially in the light of the popularity of search engines for Internet connected devices, also known as Internet of Things Search Engines (IoTSE), Internet of Everything (IoE) or Open Source Intelligence (OSINT) Search Engines such as Shodan, Censys, ZoomEye, BinaryEdge, Hunter, Greynoise, Shodan, Censys, IoTCrawler. While these tools may represent a security risk, they provide many positive and security-enhancing opportunities. They provide an overview on network security, i.e., devices connected to the Internet within the company, are useful for market research and adapting business strategies, allow to track the growing number of smart devices representing the IoT world, tracking ransomware - the number and nature of devices affected by it, and therefore allow to determine the appropriate actions to protect yourself in the light of current trends. However, almost every of these white hat-oriented objectives can be exploited by black-hatters. The popularity of IoTSE decreased a level of complexity of searching for connected devices on the internet and easy access even for novices due to the widespread popularity of step-by-step guides on how to use IoT search engine to find and gain access if insufficiently protected to webcams, routers, databases and in particular non-relational (NoSQL) databases, and other more «exotic» artifacts such as power plants, wind turbines or refrigerators. They provide service- and country- wised exposure dashboards, TOP vulnerabilities according to CVE, statistics about the authentication status, Heartbleed, BlueKeep – a vulnerability revealed in Microsoft's Remote Desktop Protocol that has become even more widely used during pandemics, port usage and the number of already compromised databases. Some of these data play a significant role for experienced and skilled attackers, making these activities even less resource-consuming by providing an overview of the ports to be used to increase the likelihood of faster access to the artifact etc.

In the past, vulnerability databases such as CVE Details were considered useful resources for monitoring the security level of a product being used. However, they are static and refer to very common vulnerabilities in the product being registered when a vulnerability is detected. Advances in ICT, including the power of the IoTSE, require the use of more advanced techniques for this purpose.

The aim of this paper is to examine both current data security research and to analyse whether "traditional" vulnerability registries provide a sufficient insight on DBMS security, or they should be rather inspected by using IoTSE-based and respective passive testing, or dynamically inspected by DBMS holders conducting an active testing. As regards the IoTSE tool, this study refers to Shodan- and Binary Edge- based vulnerable open data sources detection tool – ShoBeVODSDT - proposed in [9].

The paper is structured as follows: Section 2 provides the reader with a background, including a brief overview of data(base) security research, Section 3 gives an overview of database security threats according to the CVE Details, Section 4 provides a comparative analysis of the results extracted from the CVE database with the results obtained as a result of the application of the IoTSE-based tool. Section 5 summarizes the study, making call for "security first" principle.

# 2 Rationale of the study

Data security, and therefore database security, should be a priority for IT management as an extremely valuable asset for any organization [10-11]. Failure to comply with the requirements for security and protection of data and sensitive data in particular can lead to significant damage and losses of commercial, reputation, operational etc. nature. Recent research, however, often point out the problems associated with meeting even the most trivial requirements. A Data Breach Investigations Report [6] revealed that one of the most prominent and growing problems is the misconfiguration of DBMS. This is even more the case for NoSQL. Given that in case of NoSQL there is less focus on the security mechanism (i.e., it was not their priority), some research such as Fahd et al. [11] do not recommended to directly expose them to an open environment, where untrusted clients can directly access them. This refers to a frequently observed highly vulnerable combination of data (such as relational or document databases or cloud file storage) placed on the Internet without controls, combined with security researchers looking for them [6]. These rather undesirable combinations have been on the rise for the past few years constituting the concept of "open database". The term does not refer to data storage facilities that have been assigned the Open Database License (ODL), whereby they are knowingly made available to users to be freely used, shared, modified while maintaining the same freedom for others, thus contributing to the openness paradigm. Instead, it refers to insecure and unprotected database that can be accessed by any stakeholder despite the agreement of the data holder, which poses a serious risk. This was studied in [9], revealing that while there are databases that can be considered as open databases being accessible via Internet, the difference between NoSQL and relational database management systems (RDBMS) is not as obvious with some weak results demonstrated by SQL databases.

All in all, research aimed at identifying security threats and vulnerabilities in databases is relatively limited, with especially little research on NoSQL security, despite the vulnerabilities in NoSQL database systems is a well-known problem [9,11]. Moreover, when examining recent database research, other aspects not related to security - their performance and efficiency in certain scenarios, scalability (sharding), availability (replication), dynamism (no rigid schema) etc. appear to be more popular with much less attention paid to security [11]. An analysis of a set of key security features offered by four NoSQL systems - Redis, Cassandra, MongoDB and Neo4j [11], however, concluded that NoSQL is characterized by mostly low level of both built-in security, encryption, authentication and authorization, and auditing, while most NoSQL lack them.

The next Section is intended to provide an insight on database security provided by CVE Details – probably the most widely known registry of vulnerabilities.

# 3 CVE security vulnerability database

## 3.1 CVE scope and classification of the vulnerabilities

The CVE security vulnerability database is a free source of information providing details on disclosed cybersecurity vulnerabilities and exploits constituting a catalogue of over 172 thousand entries. These records are added to the database through a six-step process, where each interested party can contribute to the database and, if the identified vulnerability is approved, the relevant information will become part of the registry with the priority and risk assigned as a result of its discovery, where the verification is performed by the CVE participant, thereby making this list authoritative.

CVE registry divide vulnerabilities into 13 types:

1. bypass something, e.g., restriction,
2. cross-site scripting known as XSS,
3. denial of service (DoS),
4. directory traversal,
5. code execution (arbitrary code on vulnerable system),
6. gain privileges,
7. HTTP response splitting,
8. memory corruption,
9. gain / obtain information,
10. overflow,
11. cross site request forgery (CSRF),
12. file inclusion,
13. SQL injection.

## 3.2 CVE statistics of most popular databases

For the purposes of this study, the most popular databases were selected for their further analysis. The list was formed based on the results of the DB-Engines Ranking, presenting data on March 2022. In addition to the TOP-10 most popular databases CouchDB, Memcached and Cassandra were selected based on their popularity in other lists. Table I lists them along with their type and basic statistics on their vulnerabilities. The latter is retrieved from the CVE registry, where the date of the 1st and last reported vulnerability is recorded to determine if the registry provides continuous and up-to-date data, the total number of vulnerabilities reported between these dates, the most frequently reported vulnerability, and a list of 3 most popular vulnerabilities in recent 5 years. This is intended to provide some general statistics and point to current trends and whether they have changed, i.e., whether the most popular vulnerability over the years is still a key threat or the developers managed to resolve it.

Despite the undeniable popularity of NoSQL databases, relational databases remain popular, and TOP-5 consists of 4 RDBMS and MongoDB. However, at the same time, it should be noted that all the most popular relational DBMS, taking the highest places are multi-model, i.e., adapted to current trends. For example, Oracle has proven to be the most popular, using a relational DBMS as its primary database model, while secondary models include document store, graph DBMS, RDF store, and Spatial DBMS. Similarly, MySQL secondary database models are represented by document store and spatial database, as are PostgreSQL and Microsoft SQL, although the latter uses graph DBMS in addition to the above.

The highest number of discovered vulnerabilities are in MySQL, although this is the only database for which data are no longer provided, i.e., the last vulnerability was registered in 2015. MySQL is followed by PostgreSQL and IBM Db2, with Cassandra, Memcached, CouchDB, Microsoft Access, Elasticsearch and Redis reporting the fewest vulnerabilities. Although the number of revealed vulnerabilities does not necessarily mean that the level of the relevant databases is definitely higher or lower, which may depend on the popularity of these databases, users and community involvement, this suggests such an assumption. In some cases, these statistics play a decisive role in choosing a database giving the impression of a higher "security-by-design" level. At the same time, the aforementioned databases with fewer reported vulnerabilities have come under the spotlight in some of recent data leakages, with Elasticsearch dominating [13-14], from which data on unique 1.2 billion people was leaked in 2019, making this one of the largest data leaks from a single source organization in history. This also applies to perhaps the most provocative database – MongoDB, whose low security level has been widely discussed and because of which it is very often the object of IoT search engines "trainings", for which step-by-step guides are provided. Ferrari et al. [15], however, inspected compromised databases, where Redis dominated with about 30% of databases were compromised, followed by Elasticsearch (13%) and MongoDB (8%). In most cases, this was caused by misconfiguration of these databases.

For the most common and major vulnerabilities encountered over time, most of them are DoS, although code execution is also a widespread vulnerability. A database-wised analysis of the most frequently reported vulnerabilities over the past 5 years demonstrate that Code Execution is the most common and is in the TOP-3 for 11 databases, followed by overflow (7), DoS (6), bypassing something (4), gaining information (3).

**Table 1. General DB-wised statistics of their vulnerability [author, based on CVE Details]**

| **Database** | **Type of database** | **1st vulnerability registered** | **last vulnerability registered** | **Total # of vulnerabilities** | **Most popular vulnerability** | **TOP-3 vulnerabilities in 2018-2022** |
|---|---|---|---|---|---|---|
| Oracle | Relational, multi-model | 2008 | 2021 | 44 | DoS | DoS, Code Execution, Gain Information |
| MySQL | Relational, multi-model | 2001 | 2015 | 152 | DoS | - |
| Microsoft SQL Server | Relational, multi-model | 1999 | 2021 | 87 | Code Execution | Code Execution |
| PostgreSQL | Relational, multi-model | 1999 | 2022 | 134 | DoS | Code Execution, Overflow, Sql Injection |
| MongoDB | Document, multi-model | 2013 | 2022 | 38 | DoS | DoS, Code Execution, Overflow, Bypass Something |
| Redis | Key-value, multi-model | 2015 | 2021 | 23 | Overflow | Overflow, Code Execution, Memory corruption, Bypass something |
| IBM Db2 | Relational, multi-model | 2004 | 2021 | 106 | DoS | Code Execution, Overflow, Gain Information |
| Elasticsearch | Search engine, multi-model | 2018 | 2022 | 22 | Gain Information | Gain Information, DoS, Gain privilege, Code execution |
| Microsoft Access | Relational | 1999 | 2020 | 17 | Code execution | Code execution, Overflow |
| SQLite | Relational | 2009 | 2022 | 48 | DoS | Code execution, DoS, Overflow |
| Cassandra | Wide column store | 2015 | 2022 | 6 | Code execution | Code execution, DoS, Bypass Something |
| Memcached | Key-value store | 2013 | 2020 | 14 | DoS | DoS, Overflow |
| CouchDB | Document, multi-model | 2010 | 2021 | 15 | Code Execution | Code Execution, Bypass Something, Gain Privileges |

While these data are very general, Figure 1 shows data on vulnerabilities reported in 2021 and their scores, i.e., risk level (red bars indicate the highest scores or the highest number of high-risk level vulnerabilities). While for some of them data for 2022 is also provided, the purpose of this study requires to focus on 2021, when IoTSE-driven analysis by Daskevics & Nikiforova [9] took place, thereby allowing for more consistent comparison of results. At the same time, there data on Elasticsearch, Microsoft Access and Memcached for 2021 are not available. In terms of their vulnerability in 2020, however, Elasticsearch suffered most from XSS with information obtaining and DoS, Microsoft Access – code execution, where one was combined with an overflow, while for Memcached DoS was registered.

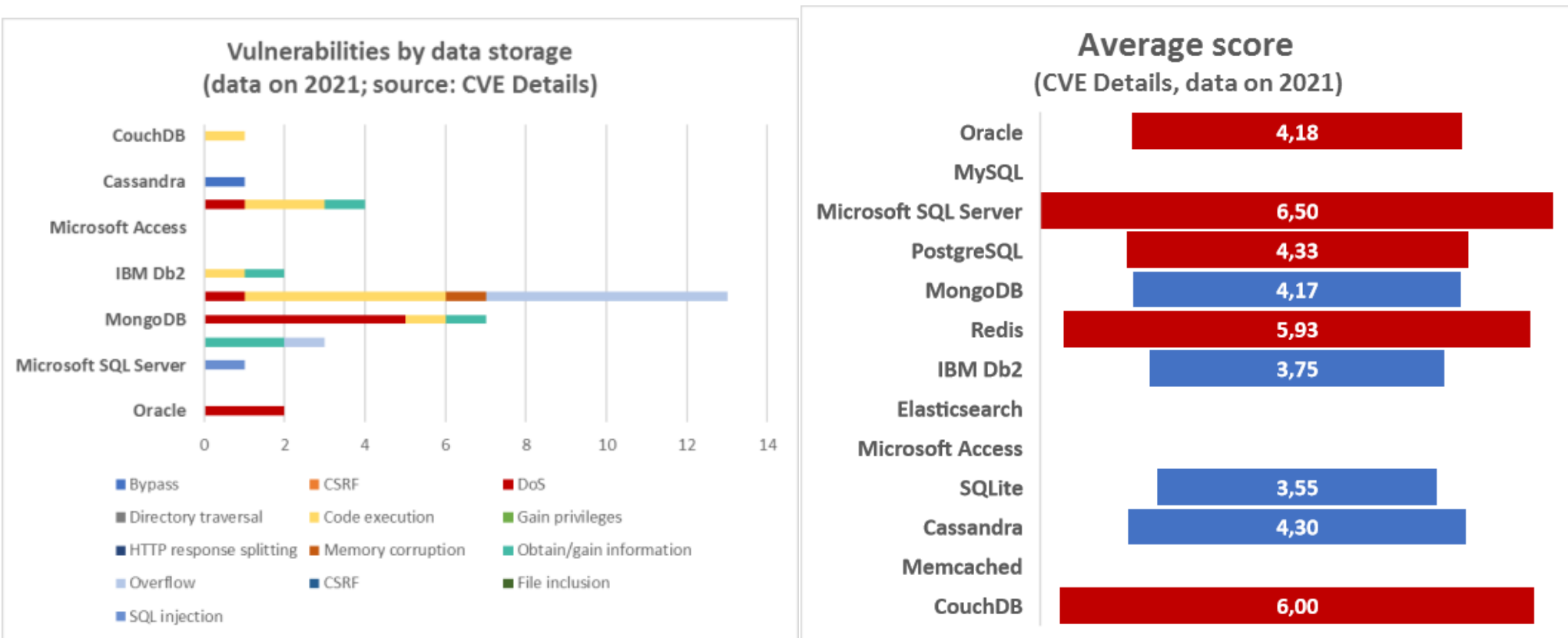

**Fig. 1. Vulnerability of data storages in 2021 and their score (based on CVE Details) [author]**

For the most popular vulnerability, the same trend is observed, i.e., 8 of 43 vulnerabilities refer to DoS, another 8 – to code execution, followed by 7 cases of overflow and 5 information leaks. Some vulnerabilities may overlap, which explains the inconsistencies in the data obtained from the registry, i.e., 31 of 43 have been supplied with corresponding detail. This, however, together with [5], according to which VulnDB has identified many thousands of vulnerabilities that were not registered in the CVE Details database, puts into question the completeness and accuracy of the CVE registry in regard to the actual state of the art.

MongoDB and Oracle have the most reported vulnerabilities, followed by Redis and PostgreSQL. Again, despite the widespread discussion about the highest risk of vulnerabilities for NoSQL compared to SQL, this is not so obvious since RDBMS are also at risk. The level of risk of registered vulnerabilities for Microsoft SQL is the highest.

Otherwise, very obvious and strong conclusions cannot be drawn from the data provided. However, it can be speculated that MongoDB is weak against DoS, but Redis against code execution and overflow. To get more supported results, this paper addresses the call made in [9] and maps the results obtained in that study to the data obtained from CVE Details.

# 4 CVE registry vs. IoTSE-based testing results

In this section, a brief overview of the results will be given, and more specifically their comparison with the results obtained by IoTSE-based tool ShoBeVODSDST that conducts a penetration testing [9]. In addition to providing both results and ranking databases by overall results, the focus will be on "Gain Information" category, as it corresponds well to the aspect inspected by ShoBeVODSDST - "managed to connect and gather sensitive data" (4th of 5 risk levels). In [9], it was expected that a correlation will be determined, allowing assumption to be made about less secure data sources.

Table 2 shows the statistics of the services under consideration, as well as their total number of vulnerabilities and the percentage of the "Gain Information" vulnerability. It should be noted that ShoBeVODSDST inspects a limited list of predefined data sources (8) with the possibility of enriching it, as its source code is publicly available.

**Table 2. CVE Details- and IoTSE- statistics on database vulnerability**

| | CVE | | | | IoTSE tool | | |
|---|---|---|---|---|---|---|---|
| **Database** | **Total # of vulnerabilities** | **Total registered** | **Ratio (Info gained/total)** | **Total DBMS found** | **# DBMS connected** | **Gathered data or compromised** | **Ratio (Info gained/ connected)** |
| Oracle | **11** | **2** | 0% | - | - | - | - |
| MySQL | - | 0 | 0% | **13452** | 0,13% | 0% | 0% |
| Microsoft SQL Server | 1 | 1 | 0% | - | | - | - |
| PostgreSQL | **5** | **3** | **67%** | **1187** | 0,17% | 0% | 0% |
| MongoDB | **13** | **7** | 14% | **177** | 8% | **79%** | 7% |
| Redis | **8** | **13** | 0% | 122 | 10% | **83%** | **83%** |
| IBM Db2 | 2 | 2 | **50%** | - | | - | - |
| Elasticsearch | - | 0 | 0% | 86 | **90%** | **27%** | **9%** |
| Microsoft Access | - | 0 | 0% | - | | - | - |
| SQLite | 2 | 1 | **50%** | - | | - | - |
| Cassandra | 1 | 1 | 0% | 7 | **14%** | 0% | 0% |
| Memcached | - | 0 | 0% | 116 | **80%** | 26% | **24%** |
| CouchDB | 1 | 1 | 0% | 14 | 0 | 0 | 0 |

According to Table 2, MySQL, the data on which is not updated by CVE, accounts more than half of all databases found on the Internet. However, the number of instances that it was able to connect to is not very high for MySQL representing 18 databases, which is similar to PostgreSQL where the number of found databases is 1187 with only 2 databases could be connected. However, there were 5 vulnerabilities in PostgreSQL registered by CVE Details, with 2 of them related to information gaining that was not found by ShoBeVODSDT. At the same time, the absolute leader in this negative trend is Memcached, where it was possible to connect to 93 of 116 databases with more than 20% of the databases, from which data have been either gathered or they were found to be already compromised. Similar results were obtained for Elasticsearch, where it was possible to connect to 90% of all databases found, and 27% of the databases are already compromised or data could be gathered from them. Similarly, CVE Details does not provide details of its vulnerabilities in 2021. MongoDB and Redis showed the worst results for both data sources, where MongoDB was inferior to data gatherings and has a large number of compromised databases according to ShoBeVODSDT and is subject to both DoS, code execution and data gatherings according to CVE Details. Redis, however, with being relatively difficult to connect to, where every 10th database is inferior to this, is characterized by a high ratio of information gatherings. According to CVE Details, both DoS, code execution, overflow, and memory corruption have been detected for it. Additionally, Oracle was one of the most frequently reported databases in CVE Details, with 10 vulnerabilities in total, while only two of them have a comprehensive description - both related to DoS.

All in all, the results in most cases are rather complimentary, and one source cannot completely replace the second. This is not only due to scope limitations of both sources - CVE Details cover some databases not covered by ShobeVODSDT, while not providing the most up-to-date information with a very limited insight on MySQL.

At the same time, there are cases when both sources refer to a security-related issues and their frequency, which can be seen as a trend and treated by users respectively taking action to secure the database that definitely do not comply with the “secure by design” principle. This refers to MongoDB, PostgreSQL and Redis. CouchDB, however, can be considered relatively secure by design, as is less affected, as evidenced by both data sources, where only one vulnerability was reported in CVE Details in 2021, while it was the only data source, to which ShoBeVODSDT was not able to connect. The latter, however, could be because CouchDB proved to be less popular, with only 14 of nearly 15 000 instances found.

# 5 Conclusions

Obviously, data security should be the top priority of any information security strategy. Failure to comply with the requirements for security and protection of data can lead to significant damage and losses of commercial, reputation, operational etc. nature [11]. However, despite the undeniable importance of data security, the current level of data security is relatively low – data leaks occur regularly, data become corrupted in many cases remaining unnoticed for IS holders. According to Risk Based Security Monthly Newsletter, 73 million records were exposed in March 2022, and 358 vulnerabilities were identified as having a public exploit that had not yet been provided with CVE IDs.

This study provided a brief insight of the current state of data security provided by CVE Details – the most widely known vulnerability registry, considering 13 databases. Although the idea of CVE Details is appealing, i.e., it supports stakeholder engagement, where each person or organization can submit a report about a detected vulnerability in the product, it is obviously not sufficiently comprehensive. It can be used to monitor the current state of vulnerabilities, but this static approach, which sometimes provides incomplete or inconsistent information even about revealed vulnerabilities, must be complemented by other more dynamic solutions. This includes not only the use of IoTSE-based tools, which, while providing valuable insight into unprotected databases seen or even accessible from outside the organization, are also insufficient.

The paper shows an obvious reality, which, however, is not always visible to the company. In other words, while this may seem surprisingly in light of current advances, the first step that still needs to be taken thinking about date security is to make sure that the database uses the basic security features: authentication, access control, authorization, auditing, data encryption and network security [11, 16-17]. Ignorance or non-awareness can have serious consequences leading to data leakages if these vulnerabilities are exploited. Data security and appropriate database configuration is not only about NoSQL, which is typically considered to be much less secured, but also about RDBMS. This study has shown that RDBMS are also relatively inferior to various types of vulnerabilities. Moreover, there is no "secure by design" database, which is not surprising since absolute security is known to be impossible. However, this does not mean that actions should not be taken to improve it. More precisely, it should be a continuous process consisting of a set of interrelated steps, sometimes referred to as "reveal-prioritize-remediate". It should be noted that 85% of breaches in 2021 were due to a human factor, with social engineering recognized as the most popular pattern [12]. The reason for this is that even in the case of highly developed and mature data and system protection mechanism (e.g., IDS), the human factor remains very difficult to control. Therefore, education and training of system users regarding digital literacy, as well as the definition, implementation and maintaining security policies and risk management strategy, must complement technical advances.

# References

1. Pevnev, V., & Kapchynskyi, S. (2018). Database security: threats and preventive measures.

2. Himeur, Y., Sohail, S. S., Bensaali, F., Amira, A., & Alazab, M. (2022). Latest Trends of Security and Privacy in Recommender Systems: A Comprehensive Review and Future Perspectives. Computers & Security, 102746.

3. Lallie, H. S., Shepherd, L. A., Nurse, J. R., Erola, A., Epiphaniou, G., Maple, C., & Belle-kens, X. (2021). Cyber security in the age of COVID-19: A timeline and analysis of cyber-crime and cyber-attacks during the pandemic. Computers & Security, 105, 102248.

4. Risk Based Security (2016) Talentbuddy.co / Talentguide.co Database Exposed, Company Reacts Swiftly, https://www.riskbasedsecurity.com/2016/05/06/talentbuddy-co-talentguide-co-database-exposed-company-reacts-swiftly/, last accessed 2022/03/31

5. Risk based security & Flashpoint (2021) 2021 Year End Report Vulnerability QuickView

6. Verizon. 2021 Data Breach Investigations Report (DBIR). 2021. 119 Pages, https://www.verizon.com/business/resources/reports/2021/2021-data-breach-investigations-report.pdf, last accessed 2022/03/31

7. Shi, F. (2020). Threat spotlight: Coronavirus-related phishing. Barracuda Networks, https://blog. barracuda. com/2020/03/26/threat-spotlight-coronavirus-related-phishing, , last accessed 2022/03/31

8. Miles B. (2022) How to minimize security risks: Follow these best practices for success, https://www.techrepublic.com/article/minimizing-security-risks-best-practices/?utm_source=email&utm_medium=referral&utm_campaign=techrepublic-news-special-offers

9. Daskevics, A., & Nikiforova, A. (2021, November). ShoBeVODSDT: Shodan and Binary Edge based vulnerable open data sources detection tool or what Internet of Things Search Engines know about you. In 2021 Second International Conference on Intelligent Data Sci-ence Technologies and Applications (IDSTA) (pp. 38-45). IEEE.

10. Li, L., Qian, K., Chen, Q., Hasan, R., & Shao, G. (2016, September). Developing hands-on labware for emerging database security. In Proceedings of the 17th Annual Conference on Information Technology Education (pp. 60-64).

11. Fahd, K., Venkatraman, S., & Hammeed, F. K. (2019). A comparative study of NoSQL sys-tem vulnerabilities with big data. International Journal of Managing Information Technology, 11(4), 1-19.

12. Verizon. 2021 Data Breach Investigations Report (DBIR). 2021. 119 Pages, https://www.verizon.com/business/resources/reports/2021/2021-data-breach-investigations-report.pdf, last accessed 2022/03/31

13. Tunggal A. (2021). The 61 Biggest Data Breaches, https://www.upguard.com/blog/biggest-data-breaches, last accessed 2022/03/31

14. Panda Security (2019), Over 1 billion people's data leaked in an unsecured server, https://www.pandasecurity.com/en/mediacenter/news/billion-consumers-data-breach-elasticsearch/,

15. Ferrari, D., Carminati, M., Polino, M., & Zanero, S. (2020, December). NoSQL Breakdown: A Large-scale Analysis of Misconfigured NoSQL Services. In Annual Computer Security Applications Conference (pp. 567-581).

16. Teimoor, R. A. (2021). A Review of Database Security Concepts, Risks, and Problems. UHD Journal of Science and Technology, 5(2), 38-46.

17. Malik, M., & Patel, T. (2016). Database security-attacks and control methods. International Journal of Information, 6(1/2), 175-183.